\documentclass[letterpaper]{article} 
\usepackage{aaai2027}  
\usepackage[hyphens]{url}  
\usepackage{graphicx} 
\usepackage{natbib}  
\usepackage{caption} 
\usepackage{amsmath,amssymb,amsfonts}
\usepackage{booktabs}                   
\usepackage{tabularx}
\usepackage{multirow}
\usepackage{tikz}
\usetikzlibrary{arrows.meta,positioning,shapes.geometric,fit,backgrounds,calc}

\newcolumntype{Y}{>{\raggedright\arraybackslash}X}

\providecommand{\Description}[1]{}

\newcommand{\runin}[1]{\par\smallskip\noindent\textbf{#1.}}

\title{Glyph: A Multi-Strategy Agentic System for Column Description and
Sensitivity-Ontology Tagging of Enterprise Data Catalogs}

\author{
    Kostia Kudriavtsev,
    Parvez Rafi,
    Sha Sundaram
}
\affiliations{
    Apple\\
    Seattle, WA and Cupertino, CA, USA\\
    k\_kudriavtsev@apple.com,
    p\_rafi@apple.com,
    shas@apple.com
}

\begin{document}
\maketitle

\begin{abstract}
Enterprise data lakes accumulate tables faster than human stewards can document or
classify them, leaving columns with missing descriptions and unassigned governance labels.
This documentation debt undermines data discovery, access control, and regulatory
compliance. We present \textbf{Glyph}, a production system that frames two coupled
problems, column \textbf{description generation} and \textbf{column type annotation for
data classification}, as cooperating LLM agents orchestrated as stateful graphs. The
\textbf{Descriptor} grounds generation in the pipeline source code that produces each
column, retrieved on demand from an enterprise GitHub via a reasoning--acting tool loop
(active Retrieval-Augmented Generation). The \textbf{Tagger} assigns labels from a governed
275-leaf \emph{Data Classification Ontology} by running three complementary strategies in
parallel (a description tagger, a line-of-business regex tagger, and a
metadata tagger backed by a fine-tuned contrastive encoder over a vector database), then fuses
their ranked outputs with Reciprocal
Rank Fusion (RRF).  We fine-tune a 6-layer MiniLM metadata encoder with an in-batch contrastive objective, lifting \emph{same-tag} retrieval on an in-distribution held-out split from NDCG@10 0.55 to 0.92 (MAP@100 $0.19\rightarrow0.90$) relative to the stock base encoder.
We report end-to-end multi-label tagging quality under a
recall-weighted $F_2$ objective across three evaluation groups, an ablation isolating each strategy and the RRF fusion, and the engineering decisions that
distinguish Glyph from prior column-type-annotation work and from commercial value/regex
sensitivity scanners: value-free and code-grounded design, per-tag provenance, and graceful
degradation. Together these make multi-agent LLM cataloging auditable and operable as a
production service.
\end{abstract}

\section{Introduction}
\label{sec:introduction}

Modern analytics organizations operate data catalogs containing tens of thousands of
tables and millions of columns. Two questions must be answered for every column before the
data can be safely used: \emph{what the column means} (documentation) and \emph{what
category of data it holds} (classification and governance). In practice both are
answered manually, if at all: descriptions are missing or stale, and sensitive-data tagging,
needed for access control and privacy compliance, lags discovery by months. The cost
is concrete: undocumented personally identifiable information (PII) is discovered reactively
\emph{after} incidents rather than proactively, and new engineers cannot reason about
datasets they did not build.

One central challenge is the \emph{scale} of dataset tagging. In practice, datasets are not
confined to a single storage technology: the same logical catalog may span Apache Iceberg
tables, relational databases (RDBMS), and Cassandra column-family stores. Consistently tagging
metadata across these heterogeneous backends, at the volume encountered in a production
environment, is intractable by hand. A single organization's production catalog spans
$\sim$3.4M columns across $\sim$98{,}000 tables, distributed over four
lines of business and the three storage backends above. At this scale, manual annotation is not
merely slow but economically infeasible: even a highly optimistic rate of one column per minute
would require roughly 30 person-years of expert effort for this catalog alone, and the
catalog drifts faster than any team can keep pace. This combination of \emph{volume} and \emph{backend
heterogeneity}, neither of which is addressed by point solutions tied to one storage
engine, is precisely what motivates an automated, learning-based approach: only a system
that generalizes across backends and improves without per-table human labeling can bring
documentation and sensitivity tagging to catalog scale.

The second task is an instance of \textbf{column type annotation (CTA)}, the assignment of a
semantic label to a column. It is a problem studied for over fifteen years, from joint graphical
models over web tables~\cite{limaye2010} through supervised deep
classifiers~\cite{hulsebos2019sherlock, zhang2020sato}, pre-trained table
encoders~\cite{suhara2022doduo, deng2021turl}, and, most recently, large language
models~\cite{korini2023cta, feuer2024archetype, kayali2024chorus}. Glyph differs from this
lineage in three ways that define its contribution. First, its target is not an open-domain
type vocabulary (DBpedia/Schema.org) but a \textbf{governed enterprise \emph{sensitivity}
ontology}: 275 leaf annotations organized by and sensitivity level,
where the dominant class (internal non-sensitive) covers roughly 72\% of columns. Second,
Glyph is \textbf{privacy-preserving by design}: it never reads cell values. The encoder sees
only the structural key \texttt{catalog.database.table.column}, and the Descriptor grounds in
the producing source code rather than the data. Reasoning about sensitive tables \emph{without ever
accessing their data} is a first-class capability, not a limitation: it satisfies enterprise privacy
and data-access policy by construction. Third, Glyph couples
heterogeneous-strategy rank fusion with an LLM-as-Judge self-refinement loop and
\textbf{per-tag provenance}, so every label carries the source, score, and reasoning behind
it, a finer-grained provenance than the rule logs of commercial scanners or the bare labels of
academic CTA systems.

Four requirements shaped the design and distinguish Glyph from a naive ``ask the model''
prototype: (1) \textbf{grounding over guessing}. Ambiguous names, such as a column
\texttt{name} that may denote either a PII-sensitive person name or a misnomer for a movie
title, must be resolved from authoritative evidence rather than parametric
guesses. (2) \textbf{Defense in depth}. Since no single signal suffices, Glyph
combines descriptions, structural metadata, historical assignments, and
naming conventions in a principled way. (3) \textbf{Measurable, self-improving quality}.
Output is verified automatically against an explicit bar and refined when it falls short.
(4) \textbf{Production discipline}: stateless horizontal scaling, deterministic decoding,
per-request authentication, structured tracing, and reproducible training.

\runin{Contributions}
(a) A two-agent architecture that separates \emph{meaning} generation (Descriptor) from
\emph{category} assignment (Tagger), connected by a shared state and ontology, and, to
our knowledge, the first published account of multi-strategy CTA targeting a governed
enterprise \emph{sensitivity} ontology with sensitivity levels. (b) A
multi-strategy tagger that fuses dense retrieval, LLM reasoning, and rule-based
signals with Reciprocal Rank Fusion (RRF)~\cite{cormack2009rrf} and a candidate-constrained LLM
re-ranker, with full per-tag provenance. (c) A contrastive fine-tuning
recipe~\cite{henderson2017} for a metadata-specific encoder, improving in-distribution
candidate recall of the metadata tagger (NDCG@10 $0.55\rightarrow0.92$, MAP@100
$0.19\rightarrow0.90$) on a per-tag, column-level held-out split of the training corpus.
(d) An end-to-end evaluation methodology built on steward-curated, per-line-of-business sets under strict
and expanding regimes, plus a strategy ablation, and a documented production stack that turns
the prototype into a reliable service.

\section{Related Work}
\label{sec:related-work}

Glyph sits at the intersection of two literatures: \textbf{semantic column type annotation}
and \textbf{enterprise metadata/governance catalogs}, which we survey in turn.

\subsection{Semantic Column Type Annotation}
\label{sec:cta}

\runin{Classical and supervised CTA}
Assigning a category label to a column was pioneered by \citet{limaye2010}, whose graphical model
jointly annotates cells, column types, and inter-column relations against a knowledge base via
\emph{collective} inference; T2K Match~\cite{ritze2015t2k} is the reference unsupervised matcher to
DBpedia. Sherlock~\cite{hulsebos2019sherlock} reframed CTA as supervised learning over character-,
word-, and statistics-level features of \emph{column values}, predicting one of 78 types at
support-weighted F1 $\approx 0.89$ on VizNet, and Sato~\cite{zhang2020sato} added table context and
a CRF for $\approx 0.92$. \citet{pham2016semantic} rank candidate labels by similarity of an unseen
column to labeled examples, and Meimei~\cite{takeoka2019meimei} is one of the few classical methods
that is explicitly \emph{multi-label} per column.

\runin{Pre-trained table encoders}
TURL~\cite{deng2021turl}, TaBERT~\cite{yin2020tabert}, and TABBIE~\cite{iida2021tabbie} pre-train
structure-aware Transformers over web tables; Doduo~\cite{suhara2022doduo} serializes an entire
table and fine-tunes a single BERT in a multi-task setup to predict column types and relations,
reaching $\approx 0.92$ micro-F1 on VizNet and remaining the canonical fine-tuned baseline.
RECA~\cite{sun2023reca} enriches a target column with \emph{related tables} retrieved by
entity-schema matching, and Watchog~\cite{miao2024watchog} pre-trains column representations
contrastively for label efficiency. Every system in this family consumes cell \emph{values}, though
Doduo and TURL read headers alongside them.

\runin{LLM-era CTA}
\citet{korini2023cta} frame CTA as text classification and prompt ChatGPT, finding that few-shot,
label-narrowing prompts approach fine-tuned baselines on SOTAB. ArcheType~\cite{feuer2024archetype}
maps free-form LLM output back to a fixed vocabulary by \emph{label remapping}, reaching near-SOTA
with open models. CHORUS~\cite{kayali2024chorus} shows one foundation model can subsume table-class
detection, CTA, and join discovery via prompt assembly, and RACOON~\cite{wei2024racoon}
retrieval-augments LLM CTA with a knowledge graph. Table-GPT~\cite{li2024tablegpt} and
TableLlama~\cite{zhang2024tablellama} instead \emph{instruction-tune} LLMs for table tasks.

\runin{How Glyph differs}
Glyph's metadata tagger (Section~\ref{sec:strategy-c}) descends from the
similarity-to-labeled-examples line but uses a fine-tuned neural encoder and ANN search over a
curated corpus rather than hand-engineered features; its regex tagger
(Section~\ref{sec:strategy-b}) is a T2K-style matcher~\cite{ritze2015t2k}; and its description
tagger (Section~\ref{sec:strategy-a}) is a retrieval-constrained, candidate-validated variant of
single-prompt LLM CTA~\cite{korini2023cta} with ArcheType-style label
grounding~\cite{feuer2024archetype}. Three differences are structural. Generation is constrained to
\emph{retrieved ontology candidates} rather than remapped post hoc. Three heterogeneous strategies
are \emph{fused} by RRF rather than one prompt being trusted, a per-column trade against the
collective inference of~\citet{limaye2010} and Sato~\cite{zhang2020sato} that we revisit in
Section~\ref{sec:discussion}. And none of these systems targets a governed \emph{sensitivity}
taxonomy from metadata alone, without cell values.

\subsection{Benchmarks}
\label{sec:benchmarks}

The dominant CTA benchmarks measure open-domain types: VizNet~\cite{hu2019viznet} supplies
the 78-type Sherlock/Sato corpus; SOTAB~\cite{korini2022sotab} labels web tables with 91
Schema.org types; GitTables~\cite{hulsebos2023gittables} annotates $\sim$1M repository-mined
tables and documents that CTA accuracy \emph{drops} on realistic relational tables;
SemTab~\cite{jimenez2019semtab} runs the annual CEA/CTA/CPA challenge against DBpedia/Wikidata.
None measures a \emph{sensitivity/data-classification} taxonomy, none is multi-label in Glyph's
sense, and all treat cell values as the primary signal even where headers are present. Because
SOTAB and GitTables do ship column headers, they remain a usable out-of-distribution probe for
Glyph's \emph{metadata-only} encoder (Section~\ref{sec:conclusion}) --- concrete near-term work.

\subsection{Industrial Catalogs and Data-Security Posture}
\label{sec:industrial}

Goods~\cite{halevy2016goods} is the canonical account of automatic enterprise metadata
enrichment at scale, crawling billions of datasets to infer schema, provenance, and summaries,
but not LLM-generated descriptions or governed sensitivity labels.
Aurum~\cite{fernandez2018aurum} and Table-Union search~\cite{nargesian2018tableunion} link
columns by content signatures for \emph{discovery}; Glyph's fine-tuned encoder plus ANN search is
the neural analogue used for \emph{tagging} rather than join discovery. Open catalogs such as
DataHub~\cite{datahub} and Amundsen~\cite{amundsen} store column tags and lineage but rely on
rule/keyword classifiers and human-authored descriptions, and Apache Atlas~\cite{atlas}
\emph{propagates} sensitivity classifications along lineage without inferring the seed label.
Commercial governance and value-scanning tools~\cite{purview, claire, presidio, clouddlp, bigid,
immuta} auto-apply classifications predominantly via regex, dictionaries, and value-pattern
matching over cell values. Glyph is the enrichment engine that produces the descriptions and seed
labels those catalogs store and propagate; by reasoning from the producing source code and
structural metadata it can classify columns whose sensitivity is \textbf{contextual rather than
format-detectable} --- a generic \texttt{recommendations} column holding a list of recommended
apps, which a value scanner sees only as high-entropy noise.

\section{Problem Formulation and the Data Classification Ontology}
\label{sec:problem}

We formalize the two problems Glyph solves, description generation and multi-label
sensitivity classification, and then describe the governed ontology that makes the
classification task fundamentally unlike open-domain column-type annotation.

\runin{Task}
Let a column be identified by its metadata key
\begin{equation}
  x = \langle \text{LOB}, \text{database}, \text{table}, \text{column}\rangle
\end{equation}
and (optionally) a natural-language description $d_x$, where LOB stands for line of business (a representation of the company's enterprise hierarchy). Glyph solves two problems: (i)
\emph{description generation}, producing $d_x$ from authoritative evidence; and (ii)
\emph{multi-label classification}~\cite{tsoumakas2007multilabel}, assigning a subset $Y_x \subseteq \mathcal{O}$ of an
ontology $\mathcal{O}$ of sensitivity annotations, each with a confidence and a justification.

\runin{The ontology}
$\mathcal{O}$ is the \emph{Data Classification Ontology}, a governed taxonomy of 275 leaf
  annotations (e.g., \texttt{USER\_NAME}, \texttt{USER\_ID}, \texttt{EMAIL\_ADDR})\footnote{Ontology
  tag names shown throughout this paper (e.g., \texttt{USER\_ID}, \texttt{EMAIL\_ADDR},
  \texttt{NON\_SENSITIVE}) are illustrative placeholders; the production Data Classification
  Ontology uses different internal codes, which are withheld for confidentiality.}
  grouped under classification categories. The categories form a
  hierarchy from generic (e.g., the \texttt{PII} category) to specific
  (e.g., the \texttt{EMAIL\_ADDR} leaf annotation). Each annotation additionally
  carries several dimensions, such as a sensitivity-level scale,
  so the ontology encodes not only \emph{what} a column is but
  \emph{how} sensitive it is and \emph{to whom}.

This makes Glyph's target fundamentally unlike the
open-domain type vocabularies of VizNet~\cite{hu2019viznet}, SOTAB~\cite{korini2022sotab},
GitTables~\cite{hulsebos2023gittables}, and SemTab~\cite{jimenez2019semtab}: the labels are
governance decisions, the task is multi-label, and the distribution is severely skewed. The
\emph{internal non-sensitive} class (\texttt{NON\_SENSITIVE}) alone accounts for $\approx 72\%$ of
columns. This imbalance is the central learning challenge and motivates both
class-balanced training and an evaluation that does not let the majority class mask
minority-class errors (Sections~\ref{sec:embedding} and~\ref{sec:evaluation}).

\runin{Why values are off-limits}
Reading cell values to classify enterprise columns is, for many of the most sensitive
datasets, precisely the action governance forbids before classification exists, a
chicken-and-egg constraint absent from value-driven CTA~\cite{hulsebos2019sherlock,
zhang2020sato, suhara2022doduo}. Glyph therefore operates on metadata and \emph{code} (which
engineers may read) rather than data, trading a signal for deployability on restricted-access
tables where Sherlock/Doduo-style scanners cannot run.

\section{Approach}
\label{sec:approach}

\subsection{System Overview}
\label{sec:overview}

\begin{figure*}[t]
\centering

{\bfseries Descriptor}\par\smallskip
\resizebox{\textwidth}{!}{%
\begin{tikzpicture}[
  font=\small, >={Stealth[round]},
  box/.style  ={draw, rounded corners, align=center, minimum height=9mm, minimum width=20mm, fill=blue!5,    inner sep=3pt},
  judge/.style={draw, rounded corners, align=center, minimum height=9mm, minimum width=20mm, fill=orange!15, inner sep=3pt},
  io/.style   ={draw, rounded corners, align=center, minimum height=9mm, minimum width=18mm, fill=black!6,   inner sep=3pt},
  arr/.style  ={->, thick},
]
\node[io]    (fs) {\texttt{fetch\_schema}};
\node[box]   (ab) [right=8mm of fs] {\texttt{analyze\_batch}\\[-2pt]{\small GitHub RAG}};
\node[judge] (gr) [right=8mm of ab] {\texttt{grade\_results}\\[-2pt]{\small LLM-as-Judge}};
\node[box]   (gd) [right=8mm of gr] {\texttt{generate\_dataset\_}\\[-2pt]\texttt{description}};
\node[io]    (fo) [right=8mm of gd] {\texttt{format\_output}};
\draw[arr] (fs)--(ab);
\draw[arr] (ab)--(gr);
\draw[arr] (gr)--(gd);
\draw[arr] (gd)--(fo);
\draw[arr, dashed] (gr) to[out=120,in=60]
      node[above, font=\small]{retry w/ feedback} (ab);
\end{tikzpicture}%
}

\vspace{5mm}

{\bfseries Tagger}\par\smallskip
\resizebox{\textwidth}{!}{%
\begin{tikzpicture}[
  font=\small, >={Stealth[round]},
  box/.style  ={draw, rounded corners, align=center, minimum height=9mm, minimum width=24mm, fill=blue!5,    inner sep=3pt},
  judge/.style={draw, rounded corners, align=center, minimum height=9mm, minimum width=24mm, fill=orange!15, inner sep=3pt},
  io/.style   ={draw, rounded corners, align=center, minimum height=9mm, minimum width=20mm, fill=black!6,   inner sep=3pt},
  arr/.style  ={->, thick},
]
\node[io]    (pc) at (0,0)       {\texttt{prepare\_columns}};
\node[box]   (A)  at (4.2,1.8)   {\texttt{llm\_desc\_tagger}\\[-2pt]{\small description-based}};
\node[box]   (B)  at (4.2,0)     {\texttt{regex\_tagger}\\[-2pt]{\small rule-based}};
\node[box]   (C)  at (4.2,-1.8)  {\texttt{llm\_metadata\_tagger}\\[-2pt]{\small metadata-only}};
\node[judge] (ev) at (8.4,1.8)   {\texttt{evaluate\_metrics}\\[-2pt]{\small LLM-as-Judge}};
\node[box]   (mg) [right=8mm of B, yshift=-3mm] {\texttt{merge\_all\_tags}\\[-2pt]{\small RRF $k=60$}};
\node[box]   (tn) [right=5mm of mg] {\texttt{tag\_normalizer}\\[-2pt]{\small tier-weighted}};
\node[io]    (fo) [right=5mm of tn] {\texttt{format\_output}};
\draw[arr] (pc)--(A);
\draw[arr] (pc)--(B);
\draw[arr] (pc)--(C);
\draw[arr] (A)--(ev);
\draw[arr, dashed] (ev) to[out=150,in=30]
      node[above, font=\small]{retry failed cols} (A);
\draw[arr] (B.east) to[out=0,in=178]   (mg.west);
\draw[arr] (C.east) to[out=25,in=200]  (mg.south west);
\draw[arr] (ev.south) to[out=-90,in=120] (mg.north west);
\draw[arr] (mg)--(tn);
\draw[arr] (tn)--(fo);
\end{tikzpicture}%
}

\caption{Glyph as two independently deployable LangGraph agents. The
\emph{Descriptor} grounds column descriptions in the producing source code via an
active-RAG tool loop, with an LLM-as-Judge refinement cycle. The \emph{Tagger} runs
three complementary strategies in parallel and fuses their ranked tag lists with
Reciprocal Rank Fusion ($k=60$), then re-scores them by sensitivity tier; a separate judge
refines the description tagger. Orange boxes
are LLM-as-Judge stages, grey boxes are I/O.}
\Description{A two-part block diagram of Glyph's two agents. The Descriptor (top) is a
left-to-right pipeline: fetch schema, batch analysis with GitHub-grounded retrieval, an
LLM-as-Judge grading step, a dataset-description synthesis step, and output formatting, with a
dashed retry-with-feedback loop from grading back to analysis. The Tagger (bottom) prepares columns and
fans out to three parallel taggers (a description tagger, a line-of-business regex
tagger, and a metadata tagger), whose ranked tag lists are merged by
Reciprocal Rank Fusion (with k equal to 60), normalized by sensitivity tier, and then formatted;
a separate LLM-as-Judge step refines the description tagger.}
\label{fig:architecture}
\end{figure*}
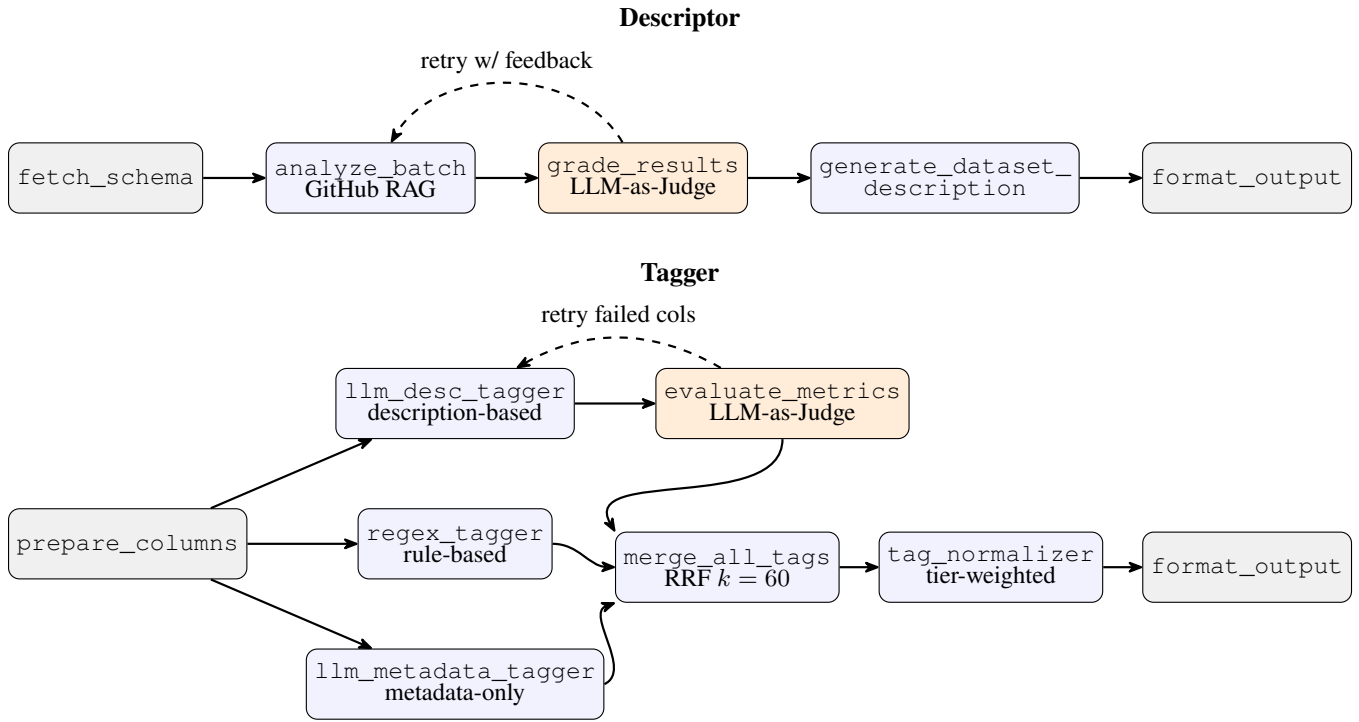

Glyph is a graph-structured agent application: each agent is a directed graph of \emph{nodes}
(pure functions over a typed state) connected by ordinary and \emph{conditional} edges, built
with LangGraph~\cite{langgraph}. Nodes are produced by a builder pattern that injects
dependencies (catalog client, LLM handles, search services) through an
\texttt{ApplicationConfig}, decoupling node logic from I/O. State is a
Pydantic~\cite{colvin2023pydantic} model, validated at every hop, and configuration is
layered in three tiers (environment defaults $\rightarrow$ per-deployment \texttt{AgentConfig}
$\rightarrow$ per-request \texttt{HyperParams}).
Fig.~\ref{fig:architecture} depicts the two agents and their nodes.

%
%

The two agents are independently deployable REST services but compose naturally: the Descriptor's generated descriptions
are the ideal input to the description tagger. Both decode at temperature 0.0
for determinism, cap token usage, and create LLM handles \emph{per request} so
model choice and credentials can vary by call.

Glyph is assembled from components whose behavior is well characterized --- active
retrieval-augmented generation, contrastive dense retrieval, multi-label classification, rank
fusion, and LLM-as-Judge critique. This is a deliberate choice for a governed sensitivity
ontology, where auditable components with known failure modes matter more than novel machinery.
The contribution is the composition: which signals are fused, how they are validated against
the ontology, and what it takes to serve them. Each mechanism is described below with its
lineage cited in place.

\subsection{The Descriptor Agent}
\label{sec:descriptor}

The Descriptor answers \emph{``what does this column mean?''} as a four-stage pipeline
(top panel of Fig.~\ref{fig:architecture}).

\runin{Schema Discovery}
\texttt{fetch\_schema} queries the enterprise data registry (the organization's internal
metadata catalog) to collect table details: column names, types, existing descriptions, etc.

\runin{Source-grounded generation (RAG)}
\texttt{analyze\_batch} is the core. To avoid the latency and inconsistency of per-column
calls, all columns are analyzed in a single LLM invocation, in batches of at most 50 columns.
The model is given the schema and a \emph{tool}, GitHub
Enterprise code search, that it may call autonomously; the evidence is \emph{code, not
data} --- the Descriptor never reads column values. This realizes active
RAG~\cite{lewis2020rag, yao2023react}, with the interleaved reasoning of chain-of-thought
prompting~\cite{wei2022cot} and in contrast to a fixed dense retriever~\cite{karpukhin2020dpr}:
the LLM decides which queries to run (a default budget
of 3 tool calls per column, capped per batch with forced finalization), reads context around
each match, and grounds descriptions in the application code that actually produces each
column.\footnote{Code search is rate-limit aware: a proactive 2\,s pause after 25 calls in a
60\,s window, plus exponential back-off (30\,s initial, doubling, up to 5 attempts).} A heuristic
confidence score is assigned to each description from observable signals (length, presence of
technical and business terms, whether the producing job was identified, and PII awareness),
capped at 1.0.

\runin{PII flagging}
A first-class objective: columns such as \texttt{user\_id} are
prefixed with explicit PII warnings, and the autograder scores PII awareness directly. Whereas
value-level scanners such as Presidio~\cite{presidio} and Cloud DLP~\cite{clouddlp} detect PII
from recognizable value \emph{formats}, Glyph reasons about sensitivity from the \emph{code and
metadata}, catching identifiers whose values are opaque hashes that format detectors miss.

\runin{Dataset description}
A \texttt{generate\_dataset\_\allowbreak description} node synthesizes a table-level summary of business
purpose from the per-column results, giving downstream consumers a self-documenting catalog
entry.

\runin{Grading and refinement}
When enabled, \texttt{grade\_results} invokes a judge model over the descriptions in
batches. The judge scores four criteria --- plausibility, specificity, completeness, and PII
awareness --- aggregated with fixed weights ($0.35$, $0.25$, $0.25$, $0.15$), and is
configured by default to be a different model family from the generator, since LLM judges
exhibit self-enhancement bias when scoring their own output~\cite{zheng2023judge}. Columns
that fall below threshold are returned to \texttt{analyze\_batch} with the
judge's per-criterion feedback; on retry the agent re-searches GitHub for the worst-scoring
columns. The loop runs up to a small fixed cap (2 passes by default).

\subsection{The Tagger Agent}
\label{sec:tagger}

The Tagger answers \emph{``what category of data is this?''} by assigning ontology tags (e.g.,
 \texttt{USER\_ID}, \texttt{EMAIL\_ADDR}) with confidence scores and reasoning. Its
defining feature is \textbf{multi-strategy parallelism with principled fusion}: three
strategies run concurrently over the columns (each fanned out into worker batches), and their
ranked tag lists are merged by a dedicated fusion node (bottom panel of
Fig.~\ref{fig:architecture}). The three are chosen so that their failure
modes are uncorrelated, allowing fusion to reconcile them.

\subsubsection{Description Tagger}
  \label{sec:strategy-a}

  The description tagger works from the \emph{meaning} of a column and is
  \emph{retrieval-constrained}: for
  each column it forms the query 
  \texttt{"\{name\}: \{description\}"} (where the description is
  the column's human-authored entry in the enterprise data registry when one exists, and the
  column name alone otherwise), retrieves the most similar ontology annotations from semantic
  search above a cosine threshold \emph{C}, and asks the LLM to select and rank the best tags
  \emph{from those candidates}, with reasoning. Because candidates are constrained to real
  ontology entries and validated afterward, hallucinated tags are rejected. This is a
  retrieval-constrained, candidate-validated variant of single-prompt LLM CTA~\cite{korini2023cta}
  with ArcheType-style label grounding~\cite{feuer2024archetype}; constraining generation to
  retrieved candidates (rather than ArcheType's post-hoc remapping) is consistent with RACOON's
  finding that retrieval improves LLM CTA~\cite{wei2024racoon}.

  The ontology annotations (for example, \texttt{EMAIL\_ADDR}: \texttt{an electronic mail address, typically written as user@domain}) are rich natural-language records, so this search
  uses a general-purpose \texttt{all-MiniLM-L6-v2}~\cite{reimers2019sbert, wang2020minilm}
  (384-dim) sentence encoder as-is.

  \runin{Grading and refinement}
  When grading is enabled, an \texttt{evaluate\_metrics} node scores the description tagger's tags with a
  combination of zero-cost deterministic metrics and a batched LLM-as-Judge~\cite{zheng2023judge};
  columns whose tags
  fall below the configured threshold are returned to \texttt{llm\_desc\_tagger} for another
  attempt, a bounded Self-Refine~\cite{madaan2023selfrefine, shinn2023reflexion} loop. The refined output
  then joins the regex and metadata taggers at fusion (Section~\ref{sec:fusion}); only the
  description tagger passes through this judge.

\subsubsection{Regex Tagger}
\label{sec:strategy-b}

The regex tagger encodes institutional knowledge by applying domain expert defined line-of-business-specific pattern rules over
column names; for example, a column \texttt{email} has a hardcoded rule that assigns the tag \texttt{EMAIL\_ADDR}. The strategy provides
high-precision anchors in the T2K matching lineage~\cite{ritze2015t2k} that complement the
learned signals.

\subsubsection{Metadata Tagger}
\label{sec:strategy-c}

The metadata tagger works from \emph{structural metadata} alone, the fully qualified key
\texttt{catalog.database.table.column}, without needing a human description, a
metadata-only descendant of value-feature classifiers~\cite{hulsebos2019sherlock} and
similarity-to-example rankers~\cite{pham2016semantic}. It is itself a RAG
system~\cite{lewis2020rag}: the metadata key is encoded with a \emph{fine-tuned}
sentence encoder (Section~\ref{sec:embedding}) and searched against an exact inner-product
(flat) vector index of verified training-corpus assignments built from human labeled corpus. Vectors are
L2-normalized so inner product equals exact cosine similarity; candidates above a configured
similarity S1 carry the tags of their nearest historical neighbors. Long, deeply nested
types are truncated to 200 characters at clean delimiter boundaries to match training-time
formatting and reduce embedding noise.

A worker then presents the retrieved candidates, with inline similarity attribution, to the LLM, which selects and ranks the final tags from these candidates. Each selected tag carries a reasoning record noting exactly which source contributed at what confidence (e.g.,
\texttt{[sources: ann=0.92]}). To our knowledge, this per-tag,
multi-source provenance is absent from prior CTA systems~\cite{hulsebos2019sherlock, zhang2020sato, suhara2022doduo,
deng2021turl, yin2020tabert, iida2021tabbie, sun2023reca, miao2024watchog, korini2023cta,
feuer2024archetype, kayali2024chorus, wei2024racoon, li2024tablegpt, zhang2024tablellama}, which
emit labels without source attribution, and finer-grained than the rule-hit logs of commercial
scanners~\cite{purview, claire, presidio, clouddlp, bigid, immuta}.

\subsubsection{Cross-Strategy Fusion with Reciprocal Rank Fusion}
\label{sec:fusion}

The three strategies produce independent ranked lists per column on incomparable score scales
(an ANN cosine, an LLM confidence, a binary rule hit). To combine
them without calibration, \texttt{merge\_all\_tags} applies \textbf{Reciprocal Rank
Fusion}~\cite{cormack2009rrf}, which needs no score calibration --- and which, unlike
self-consistency ensembling~\cite{wang2023selfconsistency}, fuses \emph{ranked lists} rather
than voting over samples. For a tag appearing at rank $r_i$ (1-indexed) in each active
source's list,
\[
  \mathrm{rrf}(t)=\sum_{i \,\in\, \text{lists containing } t}\frac{1}{k + r_i},
  \qquad k=60,
\]
normalized by the theoretical maximum $n_{\text{active}}/(k+1)$ (the score of a tag ranked
\#1 by every active source), yielding a confidence in $[0,1]$ that equals 1.0 only under
unanimous top agreement. When several strategies nominate the same tag, a fixed priority order
(\texttt{llm\_desc\_tagger} $>$ \texttt{llm\_metadata\_\allowbreak tagger} $>$ \texttt{regex\_tagger})
selects whose provenance fields (reasoning, ontology, grades) are inherited, while the
\emph{confidence} is the cross-strategy RRF score. The fused list is deduplicated by annotation
name, with the priority order above deciding which source's provenance is retained, and sorted
with a deterministic tie-break on the annotation name.

Finally, \texttt{tag\_normalizer} re-scores the fused list against the governed
\emph{sensitivity tiers} of Section~\ref{sec:problem}: it blends the RRF confidence $c$ with the
annotation's ontology tier weight $w$ as $\alpha\,c + (1-\alpha)\,w$, where untiered and higher-sensitivity annotations are unweighted ($w=1.0$) and lower-sensitivity tiers (levels 2,
3, 4) carry weights $0.75$, $0.5$, $0.25$,  respectively. Tags below a minimum normalized score are dropped and
the remainder re-sorted, so the final ranking reflects both cross-strategy agreement and
governed sensitivity level.

Table~\ref{tab:tagger-fusion-example} works this scoring through for a single column.

 \begin{table}[t]
\centering
\footnotesize
\setlength{\tabcolsep}{2pt}
\begin{tabular}{@{}lccc@{}}
\toprule
Source & \textbf{INTERNAL} & \textbf{COUNTRY} & \textbf{NON\_SENSITIVE} \\
\midrule
Regex tagger       & \checkmark &            &            \\
Description tagger &            & \checkmark &            \\
Metadata tagger    &            & \checkmark & \checkmark \\
\midrule
\textbf{RRF score} & 0.333 & 0.667 & 0.328 \\
\bottomrule
\end{tabular}
\caption{Tagger votes and fused confidence for a \texttt{country} column --- ``The 2-letter
ISO 3166-1 alpha-2 country code.'' The fused row is the normalized RRF confidence ($k=60$);
the metadata tagger ranks \texttt{COUNTRY} above \texttt{NON\_SENSITIVE}.}
\label{tab:tagger-fusion-example}

\end{table}

\subsection{Domain-Specific Embedding Fine-Tuning}
\label{sec:embedding}

The metadata tagger's ANN search (Section~\ref{sec:strategy-c}) is the one component of the Tagger that a
  general-purpose encoder cannot serve well. Its queries are metadata keys that are terse,
  abbreviation-laden, and non-linguistic, and a stock model does not know that
  \texttt{uat.core\_db.users.uid} and \texttt{prod.analytics.events.user\_id} denote the same
  concept; abbreviations, catalog prefixes, and type variants are opaque to it, and the
  off-the-shelf model leaves substantial same-tag recall unrealized in our training-corpus
  retrieval diagnostic (MAP@100 $\approx 0.19$; Table~\ref{tab:retrieval}, pre-tuning). We therefore
  \textbf{fine-tune} a compact \texttt{all-MiniLM-L6-v2} encoder (a 6-layer distillation of
  BERT~\cite{devlin2019bert}, a Transformer~\cite{vaswani2017attention} with 384 hidden units,
  $\sim$90\,MB) with an in-batch contrastive objective. This encoder is specific to the
  metadata index; the description tagger's ontology search uses a stock encoder unchanged
  (Section~\ref{sec:strategy-a}).

\runin{Training data}
A \emph{training corpus} of \texttt{metadata\_key}~$\rightarrow$~\texttt{tags} assignments is
extracted from a production system and represents already tagged and verified assignments. Pairs are
constructed within tag groups: any two columns sharing a tag form an (anchor, positive) pair.
Because a few tags (notably \texttt{NON\_SENSITIVE}, $\approx$72\%) dominate, we cap pairs per tag
(3\,000 for the reported encoder) and require a minimum group size, yielding \textbf{70\,948}
balanced training pairs.

\runin{Objective}
We use \texttt{Multiple\allowbreak Negatives\allowbreak Ranking\allowbreak Loss}~\cite{henderson2017} (scale 20, cosine similarity):
for each (anchor, positive) pair, all other positives in the batch act as negatives, pulling
same-tag columns together and pushing different-tag columns apart --- the in-batch-negatives
recipe that also underpins dense passage retrieval~\cite{karpukhin2020dpr}.

\runin{Evaluation}
A per-tag random 10\% column-level held-out query
  set, with each tag's contribution capped at 500 keys to bound memory
  (9{,}242 queries against an 83{,}229-entry corpus), is scored with an
  Information-Retrieval evaluator~\cite{reimers2019sbert, jarvelin2002ndcg} using NDCG, MRR, MAP, and
  Accuracy@k. Because the hold-out is at column granularity, columns
  from the same table may appear in both the query set and the corpus,
  so these numbers are an \emph{in-distribution upper bound} that does
  not control for within-table leakage~\cite{kaufman2012leakage}. Table-disjoint
  splitting, at (catalog, db, table) granularity, with floors of
  \(\geq\!30\) test / \(\geq\!90\) train samples per tag, is applied
  separately to the training corpus, to index the vector-search corpus; the encoder
  fine-tuning itself uses the column-level hold-out described above.
  Table~\ref{tab:retrieval} quantifies how well the encoder surfaces
  same-tag historical neighbors and therefore upper-bounds the
  candidate recall of the metadata tagger; end-to-end tagging accuracy on unseen
  tables (Section~\ref{sec:evaluation}) and out-of-distribution generalization
  (Section~\ref{sec:conclusion}) are addressed separately. The relevant set for a
  query is all same-tag corpus entries, so its size varies enormously
  by tag: huge for \texttt{NON\_SENSITIVE}, a handful for rare tags. This is why
  recall@k is small while accuracy@k stays high (\(\geq\!0.92\)
  post-tuning). MAP@100 is the most informative line because it
  rewards recovering the full same-tag pool regardless of its size:
  its \(+0.71\) gain shows the model surfaces far more of the
  relevant neighbors that the metadata tagger's LLM then re-ranks.

\begin{table}[t]
\centering
\begin{tabular}{lrrr}
\toprule
\textbf{Metric}      & \textbf{Pre-tuning} & \textbf{Post-tuning} & $\Delta$ \\
\midrule
NDCG@10     & 0.5497 & 0.9225 & $+0.3728$ \\
MRR@10      & 0.7875 & 0.9469 & $+0.1594$ \\
Accuracy@1  & 0.7138 & 0.9229 & $+0.2091$ \\
Accuracy@5  & 0.8812 & 0.9797 & $+0.0986$ \\
MAP@100     & 0.1888 & 0.9009 & $+0.7121$ \\
\bottomrule
\end{tabular}
\caption{In-distribution retrieval quality of the metadata candidate-generation encoder on a
per-tag, column-level held-out split of the training corpus, before and after contrastive fine-tuning. These are \emph{retrieval}, not \emph{tagging}, metrics.}
\label{tab:retrieval}

\end{table}

Both the encoder and the vector index are versioned and published to S3-compatible object
storage under timestamped paths; the vector index is rebuilt from the same training corpus (embedding the
\texttt{metadata\_key} \emph{without} the tag suffix to avoid train/serve skew) and loaded
read-only on container cold start. We use an exact inner-product (flat) index
because the metadata tagger runs as a stateless, ephemeral container at modest scale ($10^2$--$10^5$
vectors); scaling past $\sim$$10^6$ vectors would require a quantized or graph-based
index.

\section{End-to-End Evaluation}
\label{sec:evaluation}

Table~\ref{tab:retrieval} measures retrieval, not tagging. This section reports the metric that
is comparable \emph{in spirit} to the CTA literature: end-to-end multi-label tagging quality of
the full Tagger (all three taggers plus RRF) against curated expectations.

\begin{table*}[t]
\centering
\footnotesize
\begin{tabular}{@{}lccccccccr@{}}
\toprule
& \multicolumn{4}{c}{$F_1$} & \multicolumn{4}{c}{$F_2$} & \\
\cmidrule(lr){2-5} \cmidrule(lr){6-9}
Configuration & G1 & G2 & G3 & Overall & G1 & G2 & G3 & Overall & p50 (s) \\
\midrule
Regex tagger only               & 0.062 & 0.103 & 0.128 & 0.077 & 0.040 & 0.067 &  0.085 & 0.050 & 0.30 \\
Description tagger only, Claude Sonnet 4.5\textsuperscript{$\dagger$} & 0.122 & 0.326 & 0.411 & 0.205 & 0.250 & 0.325 & 0.590 & 0.238 & 47.31 \\
Metadata tagger only            & 0.890 & 0.969 & 0.956 & 0.912 & 0.853 & 0.957 & 0.937 & 0.880 & 40.41 \\
Fusion, Claude Sonnet 4.5                  & 0.878 & 0.951 & 0.932 & 0.897 & \textbf{0.878} & \textbf{0.961} & \textbf{0.938} & \textbf{0.890} & 57.80 \\
\bottomrule
\end{tabular}
\\[2pt]
{\footnotesize \textsuperscript{$\dagger$}The description tagger is evaluated only over
columns that carry a description; columns without one are out of scope for this strategy.}
\caption{Measured micro-averaged $F_1$ and $F_2$ per curated dataset group, overall $F_1$/$F_2$, and p50
request latency for each configuration (higher $F_1$/$F_2$ / lower latency is better). All runs cover
the same 4 evaluation files. Per-group $F_1$/$F_2$ is aggregated from per-file confusion counts. The best
$F_2$ per group is in bold.}
\label{tab:f2-bench}

\end{table*}

\runin{Harness and metrics}
Glyph ships a tag-evaluation harness that compares predicted tags against curated expectations
per column over the universe of candidate tags, computing the confusion quantities
and the derived \textbf{accuracy}, \textbf{precision}, \textbf{recall}, and the
$F_\beta$ family~\cite{manning2008ir}. While we report precision, recall, $F_{0.5}$, and $F_1$
for completeness, we adopt \textbf{$F_2$ as the primary selection metric}. The choice of $\beta$
in $F_\beta = (1+\beta^2)\,\frac{PR}{\beta^2 P + R}$ encodes a cost model, and sensitivity
tagging has a sharply \emph{asymmetric} one. A false negative, a sensitive or PII-bearing
column left \emph{untagged}, silently defeats access control and privacy compliance and is
typically discovered only reactively, after an incident. A false positive, an
over-applied tag, is cheap to remediate and dismissed by a data steward during review. We therefore select on
$F_2$ ($\beta{=}2$), which weights recall twice as heavily as precision and aligns the headline
number with the governance reality that \emph{missing} a sensitive column is far costlier than
over-tagging a benign one. 

\runin{Evaluation sets}
These sets are curated by line of business and span \textbf{multiple LOBs}, each a collection of
\texttt{\{column metadata, expected-tags\}} cases authored with data stewards. The evaluation sets
are \textbf{table-disjoint} from the training corpus of Section~\ref{sec:embedding}: none of their
1{,}905 tables appears in the 15{,}205-table vector index, though the two draw on overlapping
catalogs and databases, so no evaluated column was seen at training or indexing time.

\subsection{Per-Dataset Tagging Quality ($F_2$)}
\label{sec:per-lob}

Table~\ref{tab:f2-bench} reports measured, micro-averaged $F_2$
\emph{per curated dataset group} (Group 1--3) for four end-to-end configurations, isolating
(i) each strategy in isolation, and (ii) the full multi-strategy fusion. Per-group figures are micro-averaged from the per-file confusion counts within
each group ($F_2 = 5\,\text{TP}/(5\,\text{TP}+4\,\text{FN}+\text{FP})$); the table also reports
the overall $F_2$ and the \textbf{p50 request latency} of each configuration. We lead with $F_2$
(Section~\ref{sec:evaluation}): the ranking of configurations by $F_2$ is the ranking that minimizes
missed sensitive-data classifications.

Three findings stand out. First, \textbf{no single strategy is sufficient}: the regex tagger
alone collapses to near-zero $F_2$ across every group (it fires on almost nothing outside
its hand-written patterns), and the strongest lone strategy, the metadata tagger,
reaches only $0.880$ overall. Second, \textbf{fusion ranks above every single strategy under
$F_2$}: the best fused configuration lifts overall $F_2$ to $0.890$. This ordering
is metric-dependent, and deliberately so: the three metrics \emph{disagree on which system to
ship} --- under $F_1$ the fusion and the metadata tagger are tied, and under $F_{0.5}$ the
single strategy leads --- so the choice of $\beta$ is a deployment decision, not a cosmetic one. Third, \textbf{quality trades against latency}: fusion's $F_2$ comes at a p50 of $57.8$\,s. The description tagger is the outlier: at $47.3$\,s it spends $82\%$ of fusion's latency to reach only $0.238$, below the metadata tagger's $0.880$, so it is not a viable standalone configuration and earns its place only as an input to the fusion. The operative deployment choice is therefore metadata-only versus full fusion.

\begin{table*}[t]
\centering
\footnotesize
\setlength{\tabcolsep}{3pt}
\begin{tabularx}{\textwidth}{@{}l Y Y Y Y Y Y Y@{}}
\toprule
Dimension & \textbf{Glyph (this work)} & Sherlock &
Sato & Doduo  &
Korini \& Bizer  \\
\midrule
Input signal &
Metadata only; \textbf{no cell values} &
Cell values & Values + table topic & Serialized table values &  Values + name  \\
Learning paradigm &
Fine-tuned contrastive encoder + rules + commercial LLM &
Supervised deep neural network (DNN) & DNN + CRF & Fine-tuned PLM (multi-task)  & Zero-/few-shot LLM
 \\
Target taxonomy &
275 curated tags &
78 open types & 78 open types & 78 open types &  SOTAB types  \\
Fusion &
\textbf{RRF} & none & CRF joint & multi-task head & none  \\
Deployment &
\textbf{commercial} & research & research & research &
research   \\
Reported eval &
F1 $\approx$ 0.897 and F2 $\approx$ 0.890 & F1 $\approx$ 0.89 (VizNet) &
F1 $\approx$ 0.92 & micro-F1 $\approx$ 0.92  & F1 $\approx$ 0.85
(SOTAB)  \\
\bottomrule
\end{tabularx}
\caption{Positioning of Glyph against representative CTA systems. Systems, in column order: Sherlock~\cite{hulsebos2019sherlock},
Sato~\cite{zhang2020sato}, Doduo~\cite{suhara2022doduo} and 
\citet{korini2023cta}. }
\label{tab:positioning}

\end{table*}

\subsection{Deployed Impact: Steward Acceptance}
\label{sec:acceptance}

Glyph runs in production behind the governance workflow: human stewards adjudicate its suggestions, and each accepted or rejected \emph{column--tag} pair is written back to a data store. The training pipeline re-exports that store, re-fine-tunes the metadata encoder
(Section~\ref{sec:embedding}) --- a human-feedback loop that closes on a weekly cadence.

%
%
%
%
%
\begin{figure}[htb]
\centering
\definecolor{vizAccept}{HTML}{008300}
\definecolor{vizGrid}{HTML}{E1E0D9}
\definecolor{vizAxis}{HTML}{C3C2B7}
\definecolor{vizMuted}{HTML}{6E6C67}
\begin{tikzpicture}
  \Description{Weekly steward acceptance rate rising from 63.3 percent in week one to 99.8
  percent by week six of production operation. The vertical axis begins at 50 percent.}
  \foreach \y/\l in {0/50, 0.75/, 1.5/75, 2.25/, 3/100} {
    \draw[vizGrid,line width=0.4pt] (0,\y) -- (6.30,\y);
    \node[left=2pt,vizMuted,font=\small] at (0,\y) {\l};
  }
  \draw[vizAxis,line width=0.6pt] (0,0) -- (6.30,0);
  \node[rotate=90,anchor=south,vizMuted,font=\small] at (-0.92,1.5) {accepted \%};
  \draw[vizAccept,line width=1.1pt]
    (0,0.799) -- (1.22,1.597) -- (2.44,2.087) -- (3.66,2.920) -- (4.88,2.995) -- (6.10,2.987);
  \foreach \x/\y in {0/0.799, 1.22/1.597, 2.44/2.087, 3.66/2.920, 4.88/2.995, 6.10/2.987}
    \filldraw[vizAccept,draw=white,line width=1.1pt] (\x,\y) circle (2.0pt);
  \node[anchor=north west,font=\small] at (0.12,0.74) {63.3\%};
  \node[anchor=east,font=\small]       at (6.02,2.72) {99.8\%};
  \foreach \x/\l in {0/1, 1.22/2, 2.44/3, 3.66/4, 4.88/5, 6.10/6}
    \node[below=3pt,vizMuted,font=\small] at (\x,0) {\l};
  \node[below=1pt,vizMuted,font=\small] at (3.15,-0.52) {consecutive week};
\end{tikzpicture}
\caption{Weekly steward acceptance of Glyph's tag suggestions over six consecutive weeks of
production operation. The vertical axis begins at 50\%.}
\label{fig:acceptance}
\end{figure}
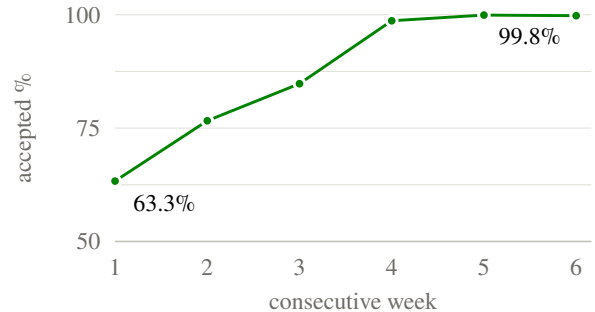

Figure~\ref{fig:acceptance} plots every steward decision to date: acceptance rises from $63.3\%$
to $99.8\%$ across $\sim24,000$ adjudicated pairs. Decision volume grows $21\times$ over the same
window and the population under review broadens alongside it, so we report the trend as an
association rather than a causal effect. Because a steward only ever
sees tags the system emitted, acceptance is a precision-like measure and is blind to false
negatives; it complements rather than replaces the recall-weighted $F_2$ of
Table~\ref{tab:f2-bench}.

\subsection{Positioning Against the CTA Literature}
\label{sec:positioning-cta}

These numbers are not directly comparable to published CTA F1 (Sherlock~\cite{hulsebos2019sherlock}, Sato~\cite{zhang2020sato}, Doduo~\cite{suhara2022doduo}, and the near-fine-tuned accuracy of ArcheType~\cite{feuer2024archetype} with open models across several
label spaces), because the taxonomy (a curated enterprise list of tags with sensitivity levels vs.\ 78--91 open-domain types without sensitivity levels), the
label cardinality (multi-label, dominated by a few specific tags), and the information available (metadata
+ code vs.\ cell values) all differ; prior-work figures are reproduced from each paper's
best-reported configuration and serve only as in-spirit reference points. 

Table~\ref{tab:positioning} positions Glyph against representative CTA systems and commercial sensitivity scanners across the dimensions that matter for enterprise deployment.

\section{Discussion and Future Work}
\label{sec:discussion}

Glyph's central bet is that \textbf{redundancy plus principled fusion} beats any single clever
signal: a rule engine that is precise but brittle, a dense retriever that generalizes but drifts,
and an LLM that reasons but hallucinates, reconciled by RRF
and an ontology validator, with an autograder converting open-ended generation into a measurable,
self-improving task. We are candid about what this paper does and does not establish.

\runin{Operational lessons}
Two design choices proved decisive in production and generalize beyond Glyph. (1)
\emph{Value-free, code-grounded design} (Section~\ref{sec:introduction}) lets Glyph run at all on
restricted-access tables that value/regex scanners~\cite{purview, claire, presidio, clouddlp, bigid, immuta}
cannot touch.
(2) \emph{Batching for cost}: analyzing a whole table's columns in one LLM call rather than one
call per column is the ratio that dominates latency and cost at catalog scale.

\runin{Future work}
Two directions to extend Glyph. (1) \emph{Collective inference.} Glyph fuses per-column
signals and ignores the inter-column and inter-tag correlation exploited since
\citet{limaye2010}, Sato~\cite{zhang2020sato}, and Meimei~\cite{takeoka2019meimei};
evaluating such joint models under our \texttt{NON\_SENSITIVE} tag dominance is a promising
direction. (2) A \emph{reinforcement-learning loop for the Tagger}: treating
tag selection as a learned policy~\cite{schulman2017ppo} optimized against a reward
model~\cite{christiano2017preferences, ouyang2022instructgpt}, so that a persistent policy learns
to rank rare tags, attacking class imbalance, and internalizes steward feedback rather than
re-deriving quality per request.

\section{Conclusion}
\label{sec:conclusion}

Glyph shows that automated, auditable data cataloging at enterprise scale is achievable by
composing well-understood techniques inside a disciplined
agentic framework, and by adapting the column-type-annotation tradition to a governed
\emph{sensitivity} ontology under an enterprise privacy constraint that forbids reading values. By
grounding descriptions in source code, fusing complementary value-free tagging signals, validating
against an authoritative ontology, and grading its own output, Glyph produces descriptions and
tags that are not only useful but \emph{explainable and auditable}: every tag carries its
provenance, confidence, and reasoning. We have been deliberate about the boundary between what is
demonstrated (strong in-distribution retrieval, a deployed multi-strategy system) and what remains
to be shown (public-benchmark generalization, judge--human agreement), and we see closing that gap
as the natural next step toward substantially reducing the manual stewardship burden of enterprise
data governance.

\bibliography{references}

\end{document}